\documentclass[12pt,a4paper]{article}

\usepackage{epsfig,graphicx}
\usepackage{amsmath,amsfonts,amssymb}
\usepackage{citesort}

\newcommand{\unit}{\leavevmode\hbox{\small1\kern-3.6pt\normalsize1}}

\def\lsim{\raise0.3ex\hbox{$\;<$\kern-0.75em\raise-1.1ex\hbox{$\sim\;$}}}
\def\gsim{\raise0.3ex\hbox{$\;>$\kern-0.75em\raise-1.1ex\hbox{$\sim\;$}}}

\begin{document}

\thispagestyle{empty}
\begin{flushright}
KUNS-1984\\
KAIST-TH 2005/13\\
FTUAM 05/11\\
IFT-UAM/CSIC-05-32\\
\vspace*{2mm}{August 29, 2005}
\end{flushright}

\begin{center}
{\Large \textbf{ More about soft terms and FCNC\\
in realistic string constructions
} }

\vspace{0.5cm}
\hspace*{-1mm}Tatsuo Kobayashi$^{a}$ and Carlos Mu\~noz$^{b,c,d}$\\[0.2cm]
{$^{a}$\textit{Department of Physics, Kyoto University, Kyoto
    606-8502, Japan.}} \\[0pt]
{$^{b}$\textit{Department of Physics, Korea Advanced Institute of
    Science and Technology,\\
Daejeon 305-701, Korea}}\\[0pt]
{$^{c}$\textit{Departamento de F\'{\i}sica Te\'{o}rica C-XI, Universidad Aut\'{o}noma de Madrid,
Cantoblanco, E-28049 Madrid, Spain}}\\[0pt]
{$^{d}$\textit{Instituto de F\'{\i}sica Te\'{o}rica C-XVI, Universidad Aut\'{o}noma de Madrid,\\
Cantoblanco, E-28049 Madrid, Spain}}\\[0pt]

\begin{abstract}
In realistic four-dimensional string constructions
the presence of anomalous $U(1)$'s is generic. In addition,
the associated Fayet-Iliopoulos contribution to the D-term can break
the
extra gauge symmetries.
As a consequence, physical particles can appear combined with other states.
We show that even if a three-generation standard-like 
model has originally flavour-independent soft 
scalar masses,
the particle mixing contribution may
generate non-universality among them.
Thus FCNC effects which were apparently absent reappear.
We also discuss the size of these contributions in an explicit model,
and how they can be suppressed.



\end{abstract}
\end{center}

\vspace*{15mm}\hspace*{3mm}
\newpage

Since the late eighties
a number of interesting four-dimensional string vacua
with particle content not far from that of the supersymmetric (SUSY)  
standard model have been found 
(see e.g. the discussion in \cite{viejos} for heterotic string models
and \cite{intersecting}  for intersecting brane models,
and references therein).
Let us recall that
the gauge groups obtained after compactification
are generically larger than the standard model gauge group, containing 
extra  $U(1)$ symmetries, $SU(3)\times SU(2)\times U(1)^n$ \cite{Kim}. 
Actually, at least 
one of these  $U(1)$'s is usually anomalous\footnote{
In \cite{Tatsuo4}, some conditions for the absence of the anomalous
$U(1)$ in heterotic string models are discussed, and classifications of models with anomalous
$U(1)$ are also attempted.}.
For example, it was found in \cite{Giedt}
that only 192 different three-generation models containing the
$SU(3)\times SU(2)\times U(1)^n$ gauge group can be constructed within the
$Z_3$ heterotic orbifold with two Wilson lines.
The matter content of 175 of them was analyzed in detail and only
7 of them turn out not to have an anomalous $U(1)$ associated.
Although the anomaly
is cancelled by the four-dimensional Green-Schwarz (GS) mechanism, it
generates a Fayet-Iliopoulos (FI) contribution to the 
D-term \cite{FayetIliopoulos}.
This effect is crucial for model building \cite{Katehou}
since some scalars 
$C$'s,
in particular $SU(3)\times SU(2)$ singlets,
acquire
large vacuum
expectation values (VEVs)
in order to cancel the FI contribution,
breaking the extra gauge symmetries, and allowing the construction
of realistic standard-like 
models \cite{Casas1,Casas2,Font}\footnote{In brane
models where several anomalous $U(1)'s$ are
usually present the FI terms do not necessarily trigger further gauge
symmetry breaking since they may vanish at the orbifold singularity.
However, the possibility of using non-vanishing FI terms for model
building is also present
and has been discussed in the literature (see e.g. \cite{standardor}).}.
In particular,
many particles, which we will refer to as $\xi$,
acquire a high mass because
of the generation of effective mass terms. These come for example
from operators of the type $C\xi\xi$.
In this way vector-like triplets and doublets and also singlets become 
heavy and disappear from the low-energy spectrum. 
The remarkable point is that the standard model matter remain massless,
surviving through certain combinations with other 
states.

The FI breaking also has an important implication in the 
flavour structure, that is, derivation of realistic 
quark/lepton mass matrices in string models.
For example, stringy 3-point couplings in heterotic 
$Z_3$ orbifold models do not lead to completely 
realistic quark/lepton mass matrices before the FI breaking\footnote{
Stringy 3-point couplings in non-prime order orbifold models 
have possibilities for realizing non-trivial quark/lepton mass 
matrices \cite{Casas:1992zt,Ko:2004ic}.
Higher dimensional operators may also be important to derive 
realistic quark/lepton mass matrices.
}.
However, after the FI breaking, the particle mixing appears 
through Yukawa couplings including scalars $C$'s with non-vanishing 
VEVs, and that can induce non-trivial quark/lepton mass matrices 
for light modes \cite{casasold}.
Detailed study on realizing quark/lepton mass matrices has been 
carried out in \cite{Abel}, showing the possibility for deriving 
realistic quark/lepton mass matrices in string models.

Generally speaking, a flavour structure leading to 
realistic fermion mass matrices may affect somehow their 
superpartner sector, that is, soft SUSY breaking sfermion masses and 
the so-called A-terms.
Thus, it is quite important to study sfermion masses and 
A-terms in each flavour structure, of course including 
the above-mentioned particle mixing scenario through 
the FI breaking.
Indeed, there are strong experimental constraints on 
flavour changing neutral currents (FCNCs).
That requires degeneracy at least between the first 
and the second generations of sfermion masses, 
unless they are sufficiently large like $O(10)$ TeV.
For example, the FCNC constraints on the Kaon system require 
\begin{eqnarray}
Re \left( (m^d_{12})^2_{LL,RR}/m^2_{ave} \right) &\leq & 
4 \times 10^{-2} \frac{m_{ave}(GeV)}{500}, \\
Im \left( (m^d_{12})^2_{LL,RR}/m^2_{ave} \right) &\leq &
3 \times 10^{-3} \frac{m_{ave}(GeV)}{500}, 
\end{eqnarray}
where $m_{ave}$ denotes the average squark mass, 
and $(m^d_{12})^2_{LL,RR}$ denotes the (1,2) entry of 
left-handed and right-handed down sector squark mass squared 
matrices in the super-CKM basis, respectively.
The latter $(m^d_{12})^2_{LL,RR}$ is given by multiplying 
the corresponding mass squared difference between the first and second 
squarks by diagonalizing matrices $(V^d_{L,R})_{12}$ of 
the left-handed and right-handed quarks.
Thus, the mass difference between the first and second squarks 
is severely constrained.
The other mass differences are constrained more weakly or there 
is no experimental constraint.

Hence, it is important to study sfermion masses in the 
particle mixing scenario through the FI breaking, because 
that happens in generic string models and moreover that has 
the possibility for deriving realistic quark/lepton mass 
matrices.
One of the important aspects in our flavour scenario is that 
the FI breaking generates D-term contributions to 
soft scalar masses
\cite{Nakano,Kawamura,Tatsuo1,Tatsuo2,Kawamura2,Higaki:2003ig,Dudas}.
Furthermore, some physical particles may appear combined with other 
states,
and this introduces another modification to scalar 
masses \cite{Tatsuo2}.
The aim of this paper is to point out 
that even if a three-generation standard-like  model 
has originally flavour-independent soft scalar masses,
the particle mixing contribution
generates non-universality among them.
Of course, depending on the size of these contributions,
the FCNC problem which was 
apparently absent in the original stringy state basis 
may reappear after the particle mixing.

Let us consider the simplest example, a model with Yukawa
superpotential
%
\begin{equation}
W=\left(\lambda_1 C_1 f
+ \lambda_2 C_2 \xi_2\right) \xi_1 \, , 
\label{massive}
\end{equation}
where $f$ denotes  a (would-be) standard model matter field,
$\xi_{1,2}$ denote two extra matter fields (triplets, doublets or singlets),
$C_{1,2}$ are the fields developing large VEVs
denoted by $\langle C_{1,2} \rangle = c_{1,2}$, and 
$\lambda_{1,2}$ are the Yukawa couplings.
In the following we will use the notation
$\tilde{c}_{1,2}\equiv \lambda_{1,2} {c}_{1,2}$.
Clearly the `old' physical particle $f$ will combine with
$\xi_{2}$. It is now straightforward to diagonalize
the scalar squared mass matrix arising from the supersymmetric mass terms in
eq.~(\ref{massive}), $|\partial W/\partial\phi_\alpha|^2$,
and the soft scalar masses, 
$m^2_\alpha |\phi_\alpha|^2$,
\begin{equation}
|\tilde c_1 f + \tilde c_2 \xi_2|^2 +
m^2_f |f|^2 
+  m^2_{\xi_2}  |\xi_2|^2 + \left(|\tilde c_1|^2 + |\tilde c_2|^2 + m^2_{\xi_1}\right) |\xi_1|^2 
\ ,
\label{massmatrisx}
\end{equation}
to
find two very massive (one of them is trivially $\xi_1$) and one light combination. 
The latter, $f'$, is obviously due to the mixing between $f$ and
$\xi_2$, 
and has a mass
\begin{equation}
m^2_{f'}= \frac{1}{2} \left\{m^2_f  + m^2_{\xi_2} +   \left(m^2_f - m^2_{\xi_2} \right) 
\frac{|\tilde c_{2}|^{2}-|\tilde c_{1}|^{2}}{|\tilde c_{2}|^{2}+|\tilde c_{1}|^{2}}\right\}
\ ,
\label{massless3}
\end{equation}
where 
we have neglected in the above formula contributions suppressed by
${O}(m_{f,\xi_2}^2/|\tilde c_{1,2}|^{2})$.
It is worth noticing here that in the case $m_{f,\xi_2}=0$, one has
$m_{f'}=0$, i.e. one recovers the limit of unbroken supersymmetry where
the combination $f'$ is massless.

Clearly, the particle mixing  generates
an additional contribution to soft scalars masses, depending on the
soft masses of extra matter $\xi$'s,  and also on 
the mixing angles $\tilde c$'s.
For the above example (\ref{massless3}), only in the particular case
when $m_f=m_{\xi_2}$ this contribution is vanishing and one 
recovers $m_{f'}=m_{f}$. 
In most of realistic models
the matter fields appear in three copies with the same quantum numbers
reproducing the three
generations of the standard model\footnote{
Recently, in \cite{Kobayashi:2004ud} a new type of models 
has been constructed, where the three generations consist of 
singlets and doublets under the $D_4$ discrete flavour symmetry. 
The $D_4$ singlets correspond to the third generation, while 
the $D_4$ doublets correspond to the first and second generations.
}, and this seems to imply
that there would be flavour-independent soft scalar masses,
since $m_{f^i}=m_{f}$, $m_{\xi_2^i}=m_{\xi_2}$, and therefore
$m_{f'^i}=m_{f'}$ as apparently deduced from the above example.
Actually, the situation 
is more involved. 

Let us consider the following explicit case:
a $Z_3$ orbifold with two Wilson lines \cite{Dixon,Wilson}
where all chiral fields appear automatically in three copies, and therefore
three-generation standard-like models have been constructed
\cite{Casas1,Casas2,Font}.
For more concreteness, one introduces non-vanishing Wilson lines for 
the first and second $T^2$'s, but not for the third $T^2$.
Hence, the degeneracy of massless spectrum on three fixed points 
of the third $T^2$ is not resolved, while degeneracy for the 
first and second $T^2$'s is resolved by non-vanishing Wilson lines.

Let us recall first that the FI breaking 
induces additional terms to 
soft scalar masses\footnote{Let us remark that
there is no additional contributions to gaugino masses and 
A-terms when Higgs fields relevant to such symmetry breaking have 
less F-term than those of dilaton and moduli fields.} due to F-terms, namely, 
the so-called D-term contribution 
\cite{Nakano,Kawamura,Tatsuo1,Tatsuo2,Kawamura2,Higaki:2003ig,Dudas}.
In particular, the presence of an anomalous $U(1)$ after
compactification
generates the dilaton-dependent 
FI term, that is, the D-term of 
the anomalous $U(1)$ is given by 
\begin{equation}
D^A=\frac{\delta^A_{GS}}{S+ S^*}+
\sum_\alpha 
(T + T^*)^{n_\alpha}
q^A_\alpha |\phi_\alpha|^2\ ,
\end{equation}
where the first term corresponds to the dilaton-dependent 
Fayet-Iliopoulos term with the GS coefficient $\delta_{GS}^A$
proportional to the value of the anomaly, and 
the second one is the usual D-term with the 
$U(1)$ charges
$q^A_\alpha$ of the 
fields $\phi_\alpha$.
Then, some of these fields (with vanishing hypercharges), let us call them
$C_\beta$,
develop large VEVs along the D-flat direction in order to cancel the FI term,
inducing the D-term contribution to the soft scalar masses of the
observable fields.
Totally, the soft scalar mass squared is given
by \cite{Tatsuo1}
\begin{equation}
m^2_\alpha = m^2_{3/2}\left\{ 1 + n_\alpha \cos ^2 \theta + 
q_\alpha^A 
\frac{\sum_\beta (T + T^*)^{n_\beta} q_\beta^A |C_\beta|^2
\left[ (6- n_\beta) \cos^2 \theta  - 5 \right]
} 
{\sum_\beta (T + T^*)^{n_\beta} (q^A_\beta)^2 |C_\beta|^2}
\right\}.
\label{totally}
\end{equation}
Here, for simplicity we have assumed that the fields $C_\beta$ 
with VEVs have no other $U(1)$ charges.
The first two terms are the usual contributions, with 
the angle $\theta$ parameterizing the direction of the goldstino in the dilaton
$S$/overall modulus $T$ field space \cite{Brignole},
and the  modular weights with typical values 
$n_{\alpha}=-1(-2)$ for untwisted(twisted) matter fields.
The third term is the D-term contribution, which is proportional to 
$U(1)$ charge $q^A_\alpha$.
Obviously, if the observable fields have vanishing $U(1)$ charges,
$q_\alpha^A=0$, this contribution is also vanishing.
However, the observable fields have usually non-vanishing charges 
in explicit models \cite{Casas1,Casas2,Font},
and the effect of this contribution must be taken into account
in the analysis.
The natural size of D-term contributions is of $O(m^2_{3/2})$, 
while in certain models \cite{Dudas} it may be enhanced.

As we can see in the above formula, the D-term contribution
generates 
an additional non-universality among soft scalar masses, 
depending on $q_\alpha^A$.
For an illustrating example, in the simplest case that 
only a single field $C$ develop a VEV in order
to cancel the FI term, the above result reduces to 
the following form:
\begin{equation}
m^2_\alpha = m^2_{3/2}\left\{ 1 + n_\alpha \cos ^2 \theta + 
\frac{q_\alpha^A}{q_C^A}\left[ (6- n_C) \cos^2 \theta  - 5 \right]
\right\}\ ,
\label{oneX}
\end{equation}
where $q_C^A$ and $n_C$ are the $U(1)$ charge and modular weight of 
the field $C$, respectively.
Notice that even in the
dilaton-dominated case ($\sin\theta =1$) the 
soft scalar masses are non-universal
\begin{equation}
m^2_\alpha = m^2_{3/2}\left( 1 - 
5 \frac{q_\alpha^A}{q_C^A}
\right)\ .
\label{dilatond}
\end{equation}

Nevertheless, for our realistic flavour scenario,
where all matter fields appear automatically in three copies (generations), 
the value of 
$q_\alpha^A$ is the same for all of them.
Also, since the three generations appear in the same twisted (untwisted) 
sector, they have the same modular weights $n_\alpha$.
Thus we have flavour-independent soft scalar masses.
Of course, we still can have
non-universal soft masses within the same generation.
This is obviously harmless from the FCNC viewpoint.
Let us then come back now to the particle mixing issue,
and study the modifications to this result.

Since all chiral fields appear automatically in three copies in this
realistic flavour scenario, the Yukawa superpotential
for the example in (\ref{massive}) must be modified as \cite{Abel}
\begin{equation}
\begin{array}{cc}
W 
& = \varepsilon' g N\ 
(\xi_1^1\ \xi_1^2\ \xi_1^3)
\left(\begin{array}{ccc}
C_{1}^1 & C_{1}^3 \varepsilon_3 & C_1^{2}  \varepsilon_3\\
C_{1}^3  \varepsilon_3 & C_{1}^2  & C_{1}^1 \varepsilon_3 \\
C_{1}^2 \varepsilon_3 & C_{1}^1 \varepsilon_3  & C_{1}^3
\end{array}\right) 
\left(\begin{array}{c}
f^{1} \\
f^{2} \\
f^{3} 
\end{array}\right)  \\
& \ +\  
\varepsilon'' g N\ 
(\xi_1^1\ \xi_1^2\ \xi_1^3) 
\left( \begin{array}{ccc}
C_{2}^1 & C_{2}^3\varepsilon_{3}  & C_2^{2}\varepsilon_{3} \\
C_{2}^3\varepsilon_3  & C_{2}^2 & C_{2}^1\varepsilon_3 \\
C_{2}^2\varepsilon_3  & C_{2}^1\varepsilon_3  & C_{2}^3
\end{array}\right)
\left(\begin{array}{c}
\xi_{2}^1 \\
\xi_{2}^2 \\
\xi_{2}^3  
\end{array}\right) 
\ ,
\end{array}
\label{mixings}
\end{equation}
%
%
where the flavour index $i$ of $f^i$ and $\xi^i_{1,2}$ 
correspond to three fixed points on the third $T^2$.
The magnitude of Yukawa couplings have been calculated explicitly 
in heterotic orbifold models 
\cite{Dixon:1986qv}.
Suppressed Yukawa couplings are obtained depending on distances among 
fixed points\footnote{
More exactly, the values of the Yukawa couplings are given by a Jacobi
theta function of the moduli fields.}.
Furthermore, the coupling selection rule allows only two types of 
combinations of fixed points on each $T^2$; 
1) all of three correspond to the same fixed point on $T^2$, 
and 2) all of three correspond to three different fixed point 
each other on $T^2$.
In the latter case, the Yukawa coupling includes 
the suppression factor $\varepsilon_i$ depending on the 
volume of the $i$-th $T^2$ as approximately
$\epsilon_i\approx 3\ e^{-\frac{2\pi}{3}T_i}$, 
where $T_i$ is the moduli parameter corresponding to the 
volume of the $i$-th $T^2$,
while the former case does not lead to such suppression factor. 
For a typical value $T_i={O}(1)$
one obtains $\epsilon_i\sim 0.1$.
In addition, $g$ in (\ref{mixings}) is the unification coupling constant
and $N$ is a quantity proportional to
the root square of the volume
of the unit cell for the $Z_3$ lattice.
Also we have $gN \sim 1$.
The factors 
$\varepsilon', \varepsilon''$ can take different values
\begin{equation}
\varepsilon' , \varepsilon'' =
1\ ,\ \varepsilon_1\ ,\  \varepsilon_2\ ,\ \varepsilon_1 \varepsilon_2\ \ ,
\label{epsilons}
\end{equation}
depending on the particular case.

Now, in order to simplify the 
analysis following \cite{Abel}, let us consider
the following VEVs for the $C_{1,2}^i$ fields\footnote{In principle 
we are allowed to do this since the cancellation of the
FI D-term only imposes $\sum_{\alpha} (T + T^*)^{n_\alpha} q_{\alpha}^{A} (|c_{\alpha}^1|^2 + 
|c_{\alpha}^2|^2 + |c_{\alpha}^3|^2)= const$,
and therefore flat directions arise.
};
\begin{eqnarray}
c_1^1\equiv c_1\ , \,\,\,\,\,\,\,\,\,\,\,\,\,\,\,\, 
c_1^2=c_1^3=0
\ ,
\nonumber
\\
c_2^1=c_2^3=0\ , \,\,\,\,\,\,\,\,\,\,\,\,\,\,\,\,
c_2^2\equiv c_2
\ .
\label{simplify}
\end{eqnarray}
Here, we expect $c_1 \sim c_2$ naturally.
Then (\ref{mixings}) gives rise to the following superpotential:
\begin{equation}
W=gN\left\{
\left(\varepsilon' C_1  f^1  
+ \varepsilon'' C_2 \varepsilon_3 \xi_2^3\right)  \xi_1^1
+
\left(\varepsilon' C_1 \varepsilon_3   f^3     
+ \varepsilon'' C_2  \xi_2^2\right)  \xi_1^2
+
\left(\varepsilon' C_1  \varepsilon_3  f^2     
+ \varepsilon'' C_2 \varepsilon_3 \xi_2^1\right)  \xi_1^3
\right\}
\ .
\label{smassive}
\end{equation}
Following the discussion for (\ref{massive}) we can deduce straightforwardly
that the masses for the three generations of physical particles $f'$
are
\begin{eqnarray}
m^2_{f'^1} &=& \frac{1}{2} \left\{m^2_f  + m^2_{\xi_2} +   \left(m^2_f - m^2_{\xi_2} \right) 
\frac{|\hat c_{2}\varepsilon_3|^{2}-|\hat c_{1}|^{2}}
{
|\hat  c_{2}\varepsilon_3|^{2}+|\hat c_{1}|^{2}}
\right\}
\ ,
\nonumber\\
m^2_{f'^2} &=& \frac{1}{2} \left\{m^2_f  + m^2_{\xi_2} +   \left(m^2_f - m^2_{\xi_2} \right) 
\frac{|\hat c_{2}|^{2}-|\hat c_{1}|^{2}}{ |\hat c_{2}|^{2}+|\hat
  c_{1}|^{2}}
\right\}
\ ,
\nonumber\\
m^2_{f'^3} &=& \frac{1}{2} \left\{m^2_f  + m^2_{\xi_2} +   \left(m^2_f - m^2_{\xi_2} \right) 
\frac{|\hat c_{2}|^{2}-|\hat c_{1}\varepsilon_3|^{2}}
{|\hat c_{2}|^{2}+|\hat
  c_{1}\varepsilon_3|^{2}
}\right\}
\ ,
\label{massless32}
\end{eqnarray}
where
\begin{equation}
\hat c_1\equiv \varepsilon' c_1 \,\,\, \ , \,\,\,\,\,\,\,\,\,
\hat c_2\equiv \varepsilon'' c_2
\ ,
\label{mixings4}
\end{equation}
and 
the soft masses $m_f=m_{f^i}$, $m_{\xi_2}=m_{\xi_2^i}$, with
$i=1,2,3$, are given
by (\ref{totally}).

Now the particle mixing contribution generates 
an additional non-universality among soft scalar masses of different
generations. Thus there may appear dangerous FCNC
effects from this lack of universality.
The question now is whether or not these contributions can be avoided 
in order not to have FCNC problems.
Notice that if we manage to have $m_f=m_{\xi_2}$, then universality
$m_{f'^i}=m_{f}$ with $i=1,2,3$
is recovered. 
In the present case of the $Z_3$ orbifold with soft masses 
given by (\ref{totally}) 
this means that the modular weights $n$ and anomalous U(1) charges $q^A$
of the fields $f$ and $\xi_2$ must be the same.
For particles in twisted sectors the modular weights are always the same,
however they can have different values of $q^A$. Thus the lack of universality
is the most general situation mainly because of the D-term 
contributions.
Nevertheless, at least an explicit example of a standard-like model can be found
\cite{Casas1,Casas2,Font}
where the relevant
couplings producing the mixings have the fields $f$ and $\xi_2$
with the same $q^A$, and therefore flavour-independent scalar masses
are obtained.
The reason for this result
is that in this model there are only two possible values
of $q^A$. Thus from  (\ref{massive}) we deduce that
$q^A_f=q^A_{\xi_2}=q^A_{C_1}=q^A_{C_2}=-q^A_{\xi_1}/2$.
Therefore, this discussion implies that a way to avoid 
the FCNC constraints is to construct string models such that the 
matter fields  $f$ and $\xi_2$ have the same $U(1)$ charges like in 
\cite{Casas1,Casas2,Font}.

Clearly, this situation does not hold in generic string model. 
For example, we can find in Appendix A of
\cite{Katehou} two other models where more possibilities for the
values of $q^A$ for the different fields are 
present\footnote{However, these models are not realistic since fields
with vanishing hypercharges cannot be found, and therefore the extra
$U(1)$ symmetries
cannot be broken. Actually those two models have a very involved
$U(1)$ combination for the hypercharge (and for the anomalous U(1)),
and we might speculate that realistic models should have simple
$U(1)$ combinations giving rise to the hypercharge and the anomalous $U(1)$, and therefore
a very limited possibilities for the values of $q^A$ for the different
fields.}.
Hence, it is important to examine whether there is another way out 
to avoid the dangerous FCNC problem in a generic situation, that is, 
the matter fields  $f$ and $\xi_2$ have different $U(1)$ charges.

Let us then discuss whether it is possible
to suppress non-degeneracy of sfermion masses due to 
the particle mixing in these orbifold models,
in the most general situation with $m_f\neq m_{\xi_2}$.
For that let us consider
three patterns for the values of 
$\hat c_{1}$, $\hat c_{2}$; 1) $\hat c_{1} \sim \hat c_{2}$, 
2) $\hat c_{1} \ll \hat c_{2}$ and 3) $\hat c_{1} \gg \hat c_{2}$, 
following \cite{Abel}.
When
$\hat c_{1} \sim \hat c_{2}$ one 
obtains from (\ref{massless32}) the following non-universality:
$m^2_{f'^1}\sim m^2_{\xi_2}$,
$m^2_{f'^3}\sim m^2_f$,
$m^2_{f'^2}\sim \frac{1}{2}(m^2_f  + m^2_{\xi_2})$.
Clearly, depending on the $U(1)$ charges of the fields
$f$ and $\xi_2$ the degree of non-universality can be important,
as discussed above.
This case may face the dangerous FCNC problem.

The model with 
$\hat c_{1} \ll \hat c_{2}$ may also be realized. For example,
this is the case when 
$\varepsilon''=1$,
$\varepsilon'=\varepsilon_1\varepsilon_2$,
and therefore using (\ref{mixings4}) one obtains
$\hat c_{2}=c_2$ and $\hat c_{1}=c_1\varepsilon_1\varepsilon_2$.
Since one expects $c_1\sim c_2$, as obtained in 
explicit models \cite{Casas1,Casas2,Font},
$\hat c_{1}$ is much smaller than $\hat c_{2}$.
As a consequence, the three generations have
$m^2_{f'^i}
\sim m^2_f$.
This result is simply understood from the fact that in this case 
the mixing between matter fields $f^i$ and $\xi^i_2$ almost vanishes 
for all of flavour indices $i$ ($i=1,2,3$), and that 
all of the three light generations approximately correspond to $f^i$.
That is not interesting for the purpose to derive realistic fermion mass 
matrices, because in this case there is no 
particle mixing effect on fermion mass matrices of three light  
generations, and these matrices even after the FI breaking are almost 
the same as those before the FI breaking.
Nevertheless, there is a subtlety in some cases, as for example when
$\varepsilon''=1$,
$\varepsilon'=\varepsilon_{1,2}$
or 
$\varepsilon''=\varepsilon_{1}$,
$\varepsilon'=\varepsilon_1\varepsilon_2$,
since now $\hat c_{1} \sim \hat c_{2}\varepsilon_{3}$ .
Then, still we have two generations, $i=2,3$ in (\ref{massless32}),
with $m^2_{f'^i}
\sim m^2_f$,
but the other has
$m^2_{f'^1}\sim \frac{1}{2}(m^2_f  + m^2_{\xi_2})$.
This may not give rise to a FCNC problem 
if we assign the first two generations
of the standard model to $f'^{2,3}$, and the third one to $f'^{1}$.
In this case, two of three light modes $f'^{2,3}$ approximately 
correspond to $f^{2,3}$, and the other light mode $f'^1$ to 
a mixture between $f^1$ and $\xi^3_2$.
That would lead to non-trivial fermion mass matrices 
because of the particle mixing.

Finally, the third pattern arises when
$\hat c_{1} \gg \hat c_{2}$, i.e. $\varepsilon' \gg \varepsilon''$.
For example with $\varepsilon''=\varepsilon_1\varepsilon_2$,
$\varepsilon'=1$,
one gets
$m^2_{f'^i}
\sim m^2_{\xi_2}$.
In this case, all of the three light generations approximately 
correspond to $\xi^i_2$, and 
there is no particle mixing effect in fermion mass matrices as the 
previous case.
Thus, this case is not interesting.
As in the previous pattern there is the subtlety
that for 
$\varepsilon''=\varepsilon_{1,2}$,
$\varepsilon'=1$
or 
$\varepsilon''=\varepsilon_{1}\varepsilon_2$,
$\varepsilon'=\varepsilon_{1,}$,
one still has
$m^2_{f'^i}
\sim m^2_{\xi_2}$ for $i=1,2$,
but 
$m^2_{f'^3}\sim \frac{1}{2}(m^2_f  + m^2_{\xi_2})$.
In this case, two of three light modes $f'^{1,2}$ 
approximately correspond to $\xi^{3,1}_2$, respectively, 
and the other light mode $f'^3$ is a mixture between 
$f^3$ and $\xi^2_2$.
If the first two generations are assigned to 
$f'^{1,2}$, there may not be the FCNC problem.

Of course, we are interested in cases leading to 
realistic fermion mass matrices.
In \cite{Abel,prepa} quark and lepton masses and mixing
angles have been
studied in the context of the $Z_3$ orbifold with mixing between
fields due to the FI breaking. In particular, interesting results have been
obtained for the structure of quark mass matrices
when
$\hat c_{1} \gg \hat c_{2}$, the first two generations correspond
to $f'^{1,2} \sim \xi^{3,1}_2$ and the third generation 
corresponds to the mixture between $f^3$ and $\xi^2_2$.
Such a case may be free from the FCNC problem as discussed above.

To summarize, we have studied sfermion masses of the 
flavour structure that all of matter fields appear originally as 
three copies and the particle mixing through the FI breaking 
leads to non-trivial quark/lepton mass matrices.
These studies are important because such particle mixing 
usually happens in generic string model, and that 
is one of the scenarios to derive realistic quark/lepton 
mass matrices.
Our result shows that although sfermion masses are 
flavour-independent in the original basis, light modes after 
particle mixing, in general, have flavour-dependent sfermion 
masses mainly due to the D-term contributions.
Therefore, this type of the flavour scenario may 
face the FCNC problem.
One way to avoid it is to construct string models such that 
mixed states corresponding to three light generations 
have the same $U(1)$ charges.
Another way out is to consider the situation that 
particle mixing effects are negligible in the first two generations 
and these fields approximately correspond to 
the original fields, while in the third  generation 
particle mixing effects are large enough to lead to 
non-trivial fermion mass matrices.
In this case, the sfermion masses between 
the first and second generations are degenerate.
Indeed, it has been shown in \cite{Abel} that this case leads to 
interesting quark mass matrices.

Here we have assumed that the scalar fields $C$'s have vanishing 
U(1) charges except the anomalous $U(1)$, and break 
only the anomalous U(1) symmetry.
However, in generic string models such scalar fields have 
non-vanishing charges other than only one $U(1)$ symmetry, 
and anomaly-free and anomalous U(1) symmetries as well as 
non-abelian symmetries are broken at the same time.
Then, complicated D-term contributions are induced as 
shown in \cite{Tatsuo1}.
However, each D-term contribution is proportional to 
U(1) charges of broken symmetries.
For such a complicated case we can repeat our analysis, and
the situation is almost the same as in the 
simple case we have studied here. That is, 
sizable non-universal sfermion masses are, in general, induced through 
the particle mixing mainly due to the D-term contributions, 
even when all of matter fields appear originally as three copies and 
sfermion masses are flavour-independent before the particle mixing.
One way to avoid the FCNC problem is to construct string models 
that mixed states have the same $U(1)$  charges for 
symmetries leading to large D-term contributions.
Another way is to consider the situation that the 
particle mixing effects are negligible for the first 
two generations, while the particle mixing in the 
third generation would lead to non-trivial 
quark/lepton mass matrices.

Let us finally remark that, even in the cases where flavour
non-universality
is present at the string scale, this not need to be necessarily a
problem \cite{Brignole,poko} 
if the effect is
substantially diluted when taking into account the flavour-independent gluino mass
contribution to the renormalization group running.
For example, the squark masses squared receive radiative corrections due to 
the gluino mass $M_{1/2}$ as $\Delta m^2 \sim 6.7 \times M^2_{1/2}$.
Thus, if the gluino mass is large sufficiently compared with 
the D-term contributions, non-universality of sfermion masses 
through the particle mixing mainly due to the 
D-term contributions would not be severely dangerous.


\noindent{\bf Acknowledgements}

\noindent 
C. Mu\~noz thanks N. Escudero for useful discussions. 
The work of T. Kobayashi was supported by 
the Grand-in-Aid for Scientific
Research \#16028211 and \#17540251, and
the Grant-in-Aid for
the 21st Century COE ``The Center for Diversity and
Universality in Physics''.
The work of C. Mu\~noz was supported 
in part by the Spanish DGI of the
MEC under Proyectos Nacionales BFM2003-01266 and FPA2003-04597;
also by the European Union under the RTN program  
MRTN-CT-2004-503369, and under the ENTApP Network of the ILIAS
project
RII3-CT-2004-506222; and also by KAIST under the Visiting Professor Program.


\end{document}